\documentclass[twocolumn,showpacs,preprintnumbers,amsmath,amssymb]{revtex4}

\usepackage{graphicx}
\usepackage{dcolumn}
\usepackage{bm}

\begin{document}

\title{Effect of depreciation of the public goods in spatial public goods games}

\author{Dong-Mei Shi$^1$ $^2$}\email{sdm21@mail.ustc.edu.cn}
\author{Yong Zhuang$^1$}
\author{Bing-Hong Wang$^1$ $^3$}
\email{bhwang@ustc.edu.cn}

\affiliation{$^{1}$Department of Modern Physics and Nonlinear
Science Center, University of Science and Technology of China, Hefei
Anhui, 230026, PR China \\$^{2}$Department of Physics, Bohai University, Jinzhou Liaoning, 121000, PR China\\
$^{3}$The Research Center for Complex System Science, University of
Shanghai for Science and Technology and Shanghai Academy of System
Science, Shanghai 200093, PR Chia
}%

\date{\today}

\begin{abstract}

In this work, depreciated effect of the public goods is considered
in the public goods games, which is realized by rescaling the
multiplication factor $r$ of each group as
$r'=r(\frac{n_{c}}{G})^{\beta}$ ($\beta\geq0$). It is assumed that
each individual enjoys the full profit of the public goods if all
the players of this group are cooperators, otherwise, the value of
the public goods is reduced to $r'$. It is found that compared with
the original version ($\beta=0$), emergence of cooperation is
remarkably promoted for $\beta>0$, and there exit optimal values of
$\beta$ inducing the best cooperation. Moreover, the optimal plat of
$\beta$ broadens as $r$ increases. Furthermore, effect of noise on
the evolution of cooperation is studied, it is presented that
variation of cooperator density with the noise is dependent of the
value of $\beta$ and $r$, and cooperation dominates over most of the
range of noise at an intermediate value of $\beta=1.0$. We study the
initial distribution of the multiplication factor at $\beta=1.0$ ,
and find that all the distributions can be described as Gauss
distribution.

\end{abstract}


\maketitle

\section{Introduction}

Evolutionary game theory \cite{IN1,IN2,IN3,IN4,IN5,IN6,IN7}is widely
applied to study the maintenance of cooperation among the selfish
individuals. The prisoners' dilemma game (PDG) by pairwise
interaction \cite{PD1,PD2,PD3} and public goods game (PGG) by group
interaction \cite{PG1} have been extensively used to investigate the
altruistic behavior. In a classical public goods game, $N$ players
are selected randomly from a large population, and each player has
two choices: cooperation and defection. A cooperator will contribute
a mount cost to the public pool, while a defector contributes
nothing. The accumulative contribution is multiplied by a factor
$r$, then redistributed equally among all the players. In a
well-mixed population, it leads to a rock-scissors-paper dynamics
when loner strategy is introduced \cite{PGG2}. Considering the
limitation of the classical game theory, some mechanisms and
theoretical supplement are proposed, such as reputation and
punishment \cite{PG2,PG3,PG4,PG5,PG6,PG7}, network reciprocity
\cite{PGG2,PG8,PG9,PG10,PG11}, voluntary participation
\cite{PG12,PG13,PG14}.

Diversity or the inhomogeneity has been studied in many works, which
mainly focuses on the network topology diversity or individual
inhomogeneity \cite{div1,div2,div3,div4,div5}. Santos $et$ $al.$
have introduced social diversity by means of heterogeneous graphs,
and concluded that cooperation is promoted by the diversity
associated with the number and the size of the public goods game, as
well as the individual contribution to each group \cite{PG11}.
Considering the profit diversity of the public goods in reality,
group diversity is introduced in the public goods game in which $r$
follows a given distribution among the population \cite{my}.
However, in real situations, the value of the public goods is not
invariable, but evolves because of the external conditions or the
reasons of themselves. In this work, we study the evolution of
cooperation in the spatial public goods games by considering the
depreciated effect of public goods. It is assumed that each
individual enjoys the full advantage of the public goods if all the
people are cooperators in a single group, otherwise, the value of
the multiplication factor of this group is reduced as a function of
$r(\frac{nc}{G})^\beta$. No qualitative change happens when altering
the initial distribution of the cooperators.

The paper is organized as follows. In section II, description of the
model is proposed in detail. Numerical simulations and the
correspondent analysis are presented in section III. Conclusions are
drawn in the Section IV.

\section{Models}

Public goods game is studied on a square lattice with periotic
boundary conditions. Each player can either cooperate or defect, and
interacts with its nearest four neighbors. Here each individual only
participates in a $SINGLE$ group. Considering the depreciated effect
of the public goods, the multiplication factor $r_{x}$ of the group
centered on individual $x$ is rescaled as

\begin{equation}
r'_{x}=r(\frac{n_{cx}}{G})^{\beta}
\end{equation}

where $r$ is the multiplication factor indicating the full profit of
the public goods of each group. $n_{cx}$ denotes the number of
cooperators in this group, and G (=5) is the group size.
$\beta\geq0$ is a tunable parameter, and when $\beta=0$, the system
returns to the original version in which $r$ is invariable and same
in each group. The payoff of player $x$ is given by

\begin{equation}
P_{x}=r'_{x}\frac{n_{cx}}{G}-s_{x},
\end{equation}

where $s_{x}$ indicates the strategy of $x$, $s_{x}$=1 for a
cooperator, and 0 for a defector.

After each time step, a player $x$ will update its strategy by
choosing a neighbor $y$ randomly from the neighborhood. We adopt the
updating rule depending on their total payoff difference,

\begin{equation}
W(s_{x}\rightarrow s_{y})=\frac{1}{1+\exp[(P_{x}-P_{y})/\kappa]},
\end{equation}

where $\kappa$ denotes the amplitude of noise level. Here we set
$\kappa=0.1$.

\section{Simulation and analysis}

Simulations are carried out for a population of $N=200\times 200$
individuals with the synchronous updating rule. Initially, the two
strategies of cooperation $(C)$ and defection $(D)$ are randomly
distributed with the equal probability of 1/2. The key quantity for
characterizing the cooperative behavior is the density of
cooperators $\rho_{c}$, which is defined as the fraction of
cooperators in the whole population. In the all simulations,
$\rho_{c}$ is obtained by averaging over the last 5000 Monte Carlo
(MC) time steps of the total 45000. Each data point results from an
average of 50 realizations.

\begin{figure}
\scalebox{0.4}[0.4]{\includegraphics{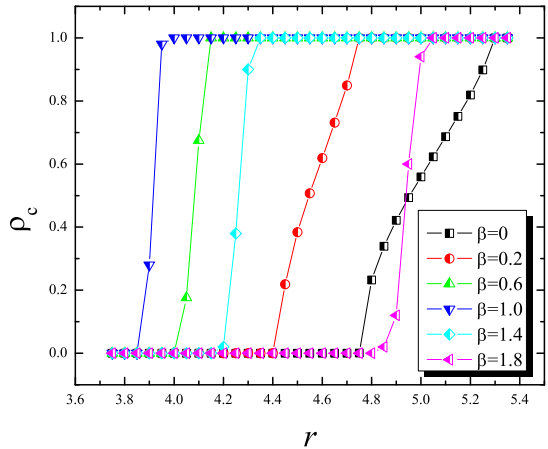}} \caption{(Color
online) Cooperator density $\rho_{c}$ as a function of
multiplication factor $r$ for different values of $\beta$.}
\end{figure}

\begin{figure}
\scalebox{0.41}[0.4]{\includegraphics{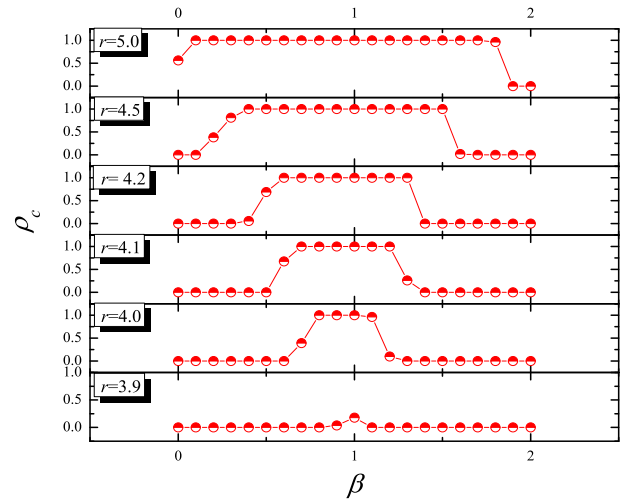}} \caption{ (Color
online) Variation of $\rho_{c}$ with $\beta$ for different values of
$r$.}
\end{figure}

Figure 1 shows the variation of $\rho_{c}$ with $r$ for different
values of $\beta$. One can see that $\rho_{c}$ increases with $r$
for each $\beta$, and emergence of cooperation is remarkably
promoted compared with the original version ($\beta$=0). Moreover,
$\beta$=1.0 induces the best promotion, which suggests that there
exit optimal values of $\beta$ for the evolution of cooperation.

Then we study the effect of $\beta$ on evolution of cooperation in
Fig. 2 for different values of $r$. It is shown that for each value
of $r$, there exit intermediate values of $\beta$ leading to the
highest cooperative level, and meanwhile, optimal plat of $\beta$ is
formed which broadens as $r$ increases. Moreover, we can also
conclude that the initial distribution of $r$ has a critical impact
on the evolution of cooperation from Fig. 2. It is shown that
$\delta$ ($\beta=0$) or polarized ($\beta\rightarrow\infty$)
distribution has an adverse effect on evolution of cooperation,
which is consistent with the results in the previous work.

Before we further study, it is necessary to explain the results
gained above. It is known that group diversity can promote
cooperation in the public goods games in which the multiplication
factor $r$ follows a given distribution \cite{my}. In our model,
diversity of group forms when $\beta>0$ according to equation (1),
and meanwhile, the diversity is not fixed, but evolves with the time
step as cooperation evolves. The spatio-temporal diversity
contributes to the spread of cooperation, and thereby better
promotes the cooperative level than the fixed spatial diversity.
Because in the fixed distribution of diversity, the players in
disadvantageous group with very small values of $r$ will always give
up cooperating (see reference \cite{my}).

\begin{figure}
\scalebox{0.4}[0.42]{\includegraphics{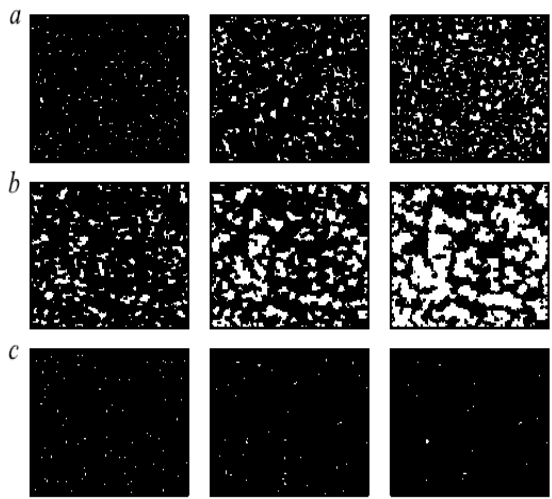}} \caption{ Time
evolution of spatial distributions of cooperators (white) and
defectors (black) at $r$=4.5 for ($a$) $\beta$=0.2, (b) $\beta$=1.0,
(c) $\beta$=1.8. The initial distribution of cooperators is uniform
distribution for each condition.}
\end{figure}

Figure 3 shows the time evolution of cooperation for different
values of $\beta$ at $r$=4.5. It is found that for small $\beta$
($\beta$=0.2), cooperative clusters are formed, and sustained as
small sizes; when $\beta$ is moderate, large clusters are formed,
and are spreading over all the population as $t$ evolves. While for
the large values of $\beta$, cooperative clusters can't be
sustained, and gradually exploited by the defectors with the time
step.

\begin{figure}
\scalebox{0.4}[0.42]{\includegraphics{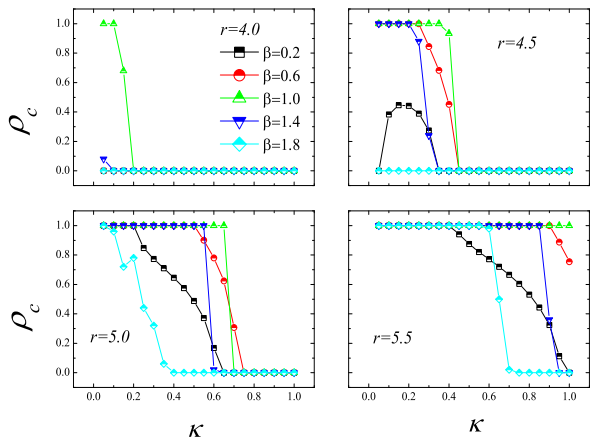}} \caption{ (Color
online) Cooperator density $\rho_{c}$ as a function of noise
amplitude $\kappa$ for different values of $\beta$ at $r$=4.0, 4.5,
5.0, 5.5 respectively.}
\end{figure}

\begin{figure}
\scalebox{0.4}[0.42]{\includegraphics{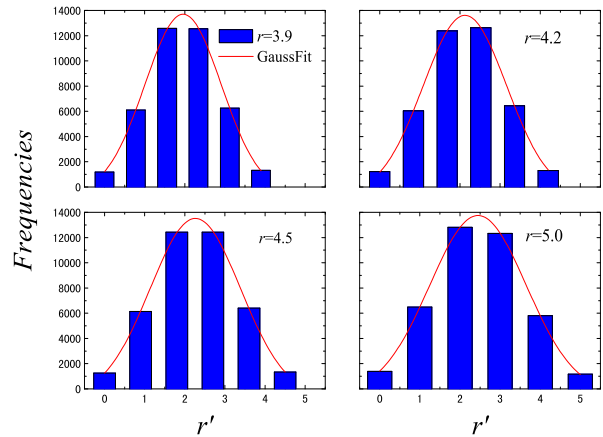}} \caption{ (Color
online) Initial distribution of $r'$ at $\beta=1.0$ for different
values of $r$. Here $\kappa$=0.1. }
\end{figure}

We further investigate the impact of noise on the evolution of
cooperation in Fig. 4 for different values of $r$. Apparently, the
variation tendency of cooperator density $\rho_{c}$ with $\kappa$
not only depends on the value of $r$, but also has the relation with
the value of $\beta$: when $r$ is too small or too large, $\rho_{c}$
decreases with $\kappa$ for each $\beta$, while for a moderate value
of $r$, $\rho_{c}$ plays nonmonotonic behavior at some value of
$\beta$ (e.g., see $r$=4.5 at $\beta$=0.2). It should be also noted
that cooperation almost dominates at an intermediate value of
$\beta$=1.0 for each $r$.

Given that the initial distribution of the group diversity plays an
important part on the evolution of cooperation, we study the initial
distribution of multiplication factor $r'$ for different values of
$r$ at $\beta=1.0$ (see Fig. 5) in order to figure out why
$\beta$=1.0 has such an advantageous status in Fig. 4. It is
surprisingly found that all the initial distribution can be
described as Gauss distribution, which implies that Gauss
distribution has an positive effect on the evolution of cooperation.

\section{Conclusions}

In this work, evolution of cooperation was investigated in the
spatial public goods games by considering the effect of depreciation
of the public goods. We assumed that each individual enjoys the full
profit of the public goods if all the players of this group are
cooperators, otherwise, the value of public goods is depreciated,
which is realized by rescaling the multiplication factor $r$ of this
group to $r(\frac{n_{cx}}{G})^{\beta}$. So that spatial diversity of
the multiplication factor is formed among the population, and
meanwhile, evolves with the time step as the cooperation evolves. We
found that compared with the original version for $\beta=0$,
cooperative level is largely improved when $\beta>0$, and there
exist moderate values of $\beta$ leading to a best cooperation.
Moreover, optimal flat of $\beta$ are formed, and broadens as $r$
increases. The facilitation of cooperation should be attributed to
the spatio-temporal diversity of the multiplication factor caused by
the depreciated effect of the public goods: spatial diversity
stimulates the emergence of cooperation, whereas temporal diversity
of $r$ contributes to the spread of cooperation. Furthermore, noise
impact on the evolution of cooperation was studied, it was proved
that the variation of cooperation density not only depends on the
values of $r$, but also has a close relation with the values of
$\beta$. We also noted that there exits an intermediate value of
$\beta=1.0$ inducing the best cooperation over most of the range of
noise. In order to explain this phenomenon, initial distribution of
multiplication factor at $\beta=1.0$ is studied. It was shown that
all the distribution can be depicted by the Gauss distribution,
which suggested that Gauss distribution has a positive effect on the
evolution of cooperation.

Our work reiterates the importance of the diversity on the evolution
of cooperation, and meanwhile, that Gauss distribution has such an
advantageous status may give us some inspiration to study the
cooperation in a different way.

This work is funded by the National Basic Research Program of China
(973 Program No. 2006CB705500), the National Important Research
Project: (Study on emergency management for non-conventional
happened thunderbolts, Grant No. 91024026), the National Natural
Science Foundation of China (Grant Nos. 10975126, 10635040), the
Specialized Research Fund for the Doctoral Program of Higher
Education of China (Grant No. 20093402110032).

\end{document}